\documentclass[conference]{IEEEtran}
\IEEEoverridecommandlockouts
\usepackage{cite}
\usepackage{amsmath,amssymb,amsfonts}
\usepackage{algorithmic}
\usepackage{graphicx}
\usepackage{textcomp}
\usepackage{xcolor}
\usepackage{multirow, makecell}
\usepackage{textcmds}
\usepackage{threeparttable}
\usepackage{scrextend}
\usepackage{cleveref}
\usepackage{url}
\usepackage{booktabs}
\usepackage{color,colortbl}
\def\BibTeX{{\rm B\kern-.05em{\sc i\kern-.025em b}\kern-.08em
    T\kern-.1667em\lower.7ex\hbox{E}\kern-.125emX}}
\definecolor{grayone}{gray}{.8}

\newcommand{\comment}[1]{}
\newcommand{\tvar}[1]{\mathit{#1}}
\usepackage{enumitem}
\setitemize[0]{leftmargin=15pt}
\usepackage{balance}

\usepackage[font=small,skip=1pt]{caption}
\newcommand{\distance}{1pt}
\usepackage[normalem]{ulem}

\begin{document}
\title{SeqMobile: An Efficient Sequence-Based Malware Detection System Using RNN on Mobile Devices}

\author{\IEEEauthorblockN{
		Ruitao Feng\IEEEauthorrefmark{1}\IEEEauthorrefmark{2}\IEEEauthorrefmark{3},
		Jing Qiang Lim\IEEEauthorrefmark{2}\IEEEauthorrefmark{3},
		Sen Chen\IEEEauthorrefmark{4}\IEEEauthorrefmark{3},
		Shang-Wei Lin\IEEEauthorrefmark{3}, and
		Yang Liu\IEEEauthorrefmark{3}
		\IEEEcompsocitemizethanks{
		\IEEEcompsocthanksitem \IEEEauthorrefmark{1}Ruitao Feng is the corresponding author.
		\IEEEcompsocthanksitem \IEEEauthorrefmark{2}These two authors contributed equally to this paper.
		}}
		\IEEEauthorblockA{\IEEEauthorrefmark{3}School of Computer Science and Engineering, Nanyang Technological University, Singapore\\
		\IEEEauthorrefmark{4}College of Intelligence and Computing, Tianjin University, China}
}

\maketitle

\begin{abstract}
With the proliferation of Android malware, the demand for an effective and efficient malware detection system is on the rise. The existing device-end learning based solutions tend to extract limited syntax features, such as permissions and API calls, to meet a certain time constraint of mobile devices. 
However, unlike sequence-based features, syntax features lack the semantics which can represent the potential malicious behaviors and further result in more robust model with high accuracy for malware detection.

In this paper, we propose an efficient Android malware detection system, named SeqMobile, which adopts behavior-based sequence features and leverages customized deep neural networks on mobile devices instead of the server end. Different from the traditional sequence-based approaches on server end, to meet the performance demand on mobile devices, SeqMobile accepts three effective performance optimization methods to reduce the time of feature extraction and prediction. To evaluate the effectiveness and efficiency of our system, we conduct experiments from the following aspects 1) the detection accuracy of different recurrent neural networks (RNN); 2) the feature extraction performance on different mobile devices, and 3) the detection accuracy and prediction time cost of different sequence lengths. The results unveil that SeqMobile can effectively detect malware with high accuracy. Moreover, our performance optimization methods have proven to improve the performance of training and prediction by at least twofold.
Additionally, to discover the potential performance optimization from the state-of-the-art TensorFlow model optimization toolkit for our sequence-based approach, we also provide an evaluation on the toolkit, which can serve as a guidance for other systems leveraging on sequence-based learning approach.
Overall, we conclude that our sequence-based approach, together with our performance optimization methods, enable us to efficiently detect malware under the performance demands of mobile devices.
\end{abstract}

\begin{IEEEkeywords}
Malware detection, Sequence feature, Deep neural network, Mobile platform, Performance
\end{IEEEkeywords}

\section{Introduction}
Smartphones have revolutionized our lives for the better in some ways. Besides calling and sending text messages, people are using these devices to watch movies, perform banking transactions~\cite{chen2018mobile,chen2019ausera}, read the news, etc. It is undeniable that these smartphones have yielded many benefits for society, allowing millions of people to stay connected through the Internet. Consequently, it has also drawn the attention of malware authors to disseminate their malware on the application markets (e.g., Google Play Store)~\cite{tang2019large}.
However, unlike the App Store for Apple iOS, the protocol for uploading an application on the Android application market is not that stringent. Therefore, there is a demand for an effective malware detection system running on the device to address the above security problem.

Currently, majority of the machine learning-based malware detection systems performed their analysis on the server side~\cite{chen2016stormdroid,chen2018automated}, until Drebin~\cite{arp2014drebin}, which is a lightweight method for detection of Android malware that enables identifying malicious applications using machine learning directly on the smartphone. Recently, researchers have been looking at ways to implement effective solutions with deep learning techniques on the device-end. Due to the performance limitations on mobile devices, researchers tend to extract limited syntax features to meet certain time constraints~\cite{feng2019performance, feng2020performancesensitive}. 
{Although using syntax features without semantics (e.g., order, position) has achieved a relatively high accuracy, it will consequently fail to maintain a more robust detection system by providing necessary information to represent certain malicious behaviors.}
Therefore, there is a demand to research into ways to effectively represent meaningful and robust features such that it contains more semantics on the malicious behaviors with limited performance overhead on mobile devices.
Unlike syntax feature based learning approaches, the feature input for sequence-based learning approaches provides not only the existence of each determined syntax feature, but also represents the semantics corresponding to certain behavior patterns~\cite{mclaughlin2017deep, robinnix2017, opcodeseq2014}. 
However, due to the complexity of sequence-based feature, traditional server-end sequence-based learning approaches failed to satisfy the demand of run-time performance when it comes to detection on mobile devices.
To address these problems, we intend to provide an efficient sequence-based malware detection system using deep learning on mobile devices.

In this paper, we propose SeqMobile, which adopts behavior-based sequence features and customized deep neural networks to provide an effective and efficient malware detection service on Android devices. To enhance the performance of SeqMobile, we propose a series of performance optimization methods that can effectively reduce the training and prediction time for sequence-based approaches.
In the experiments, we first summarize and propose {8} feature categories (e.g., combination of permissions, intent filters, API sequence and intent sequence) and investigate their corresponding performance with different deep neural networks. We then perform an evaluation using accuracy and prediction time as metrics to decide on a suitable network configuration. After that, we accept the pre-trained model that yields the best results and deploy it onto Android devices. To ensure that our pre-trained models are compatible with Android device, we convert our pre-trained models into lightweight TensorFlow Lite models.
In our proposed system, we prioritize time cost over accuracy such that lower-end devices can choose to trade off less than 1\% of the classification accuracy for lower prediction time. Overall, through our performance optimization methods, SeqMobile can achieve a relatively higher classification accuracy (i.e., 97.85\%) as well as lower feature extraction and prediction time cost (i.e., $<$5s).

In this paper, we make the following contributions:
\begin{itemize}
    \item We propose an efficient sequence-based malware detection system, which adopts behavior-based sequence feature and customized deep neural network to provide an effective and efficient malware detection service on Android devices.
    \item We present a systematic approach to directly extract the semantic feature sequence, which can provide information of certain malicious behaviors, from binary files under a certain time constraint. Thereby, achieving a relatively higher classification accuracy (i.e., 97.85\%). 
    \item We propose a method to remove repetitive elements in sequences and further evaluate how it can affect the overall performance of our malware detection system. Results has shown that our removal method significantly enhances the training and prediction performance with insignificant effects on the accuracy. To our best knowledge, this is the first comprehensive study on how removing repetitive elements in sequences can affect training and prediction performance in sequence-based learning approach on mobile devices.
    \item We conduct an evaluation on the state-of-the-art mobile-end model optimization toolkit provided by TensorFlow for our proposed sequence-based learning approach. The evaluation results can serve as a guidance for other mobile-end sequence-based learning approaches.
\end{itemize}

\begin{figure*}
\centering
\includegraphics[width=0.8\textwidth]{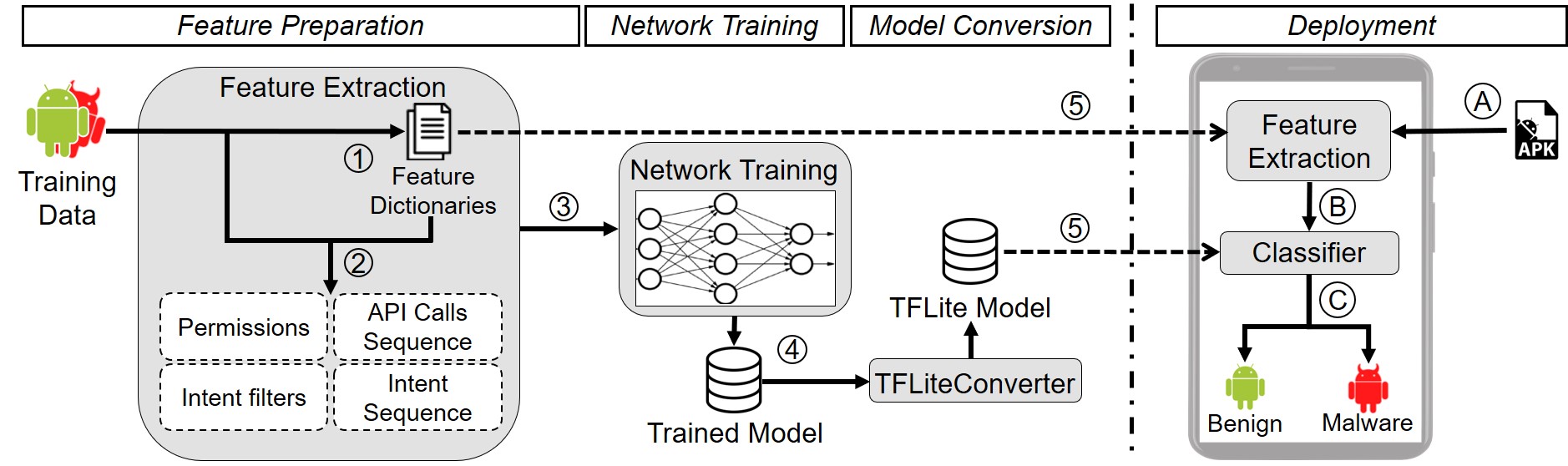}
\caption{Overview of SeqMobile}
\label{fig:OverviewDetailed}
\end{figure*}
\section{Background}\label{background}

\subsection{Sequence Representation of Application Behavior}
{Mostly, Android applications provide their functionalities with basic behaviors that are represented using permissions, intent, API calls and etc. However, the analysis of Android malware shows that there are high risks that those basic behaviors inside applications may be accepted as a part of malicious functionalities. For example, considering a spyware, no matter how much it hides its malicious functionality, there will still be necessary basic behaviors existing to access the private information from those devices. Thus, a semantic representation of the basic behaviors, like sequence-based feature, will be beneficial in providing the corresponding potential malicious information unlike the traditional syntax feature in learning approaches.}

\subsection{Native Code Implementation}
{Considering the architecture of an Android system, the functionalities in Java implementation will be executed on Dalvik Virtual Machine. Different from traditional operating systems, when facing computationally intensive tasks, significant performance problems may occur on mobile devices. To meet certain performance criteria on Android devices, developers often investigate the bottleneck of their code and attempt improve the performance by re-implementing it using native code (C and C++), which can then be invoked through the Java Native Interface (JNI){~\cite{jniframework}}. As a result, Google recommends developers to perform the heavy operations on the native-end and return the results to the Java-end through JNI.}

\subsection{Model Quantization}
{Popular deep learning frameworks such as TensorFlow~\cite{tensorflow} has provided a state-of-the-art model optimization toolkit{~\cite{tflitemodeloptimizer}} to help optimized pre-trained models on mobile devices. Quantization is one of the features provided in the toolkit that can reduce the model size and prediction time by reducing the precision of the parameters inside the model (e.g., float32 to float16).}

\subsection{Static and Dynamic RNN}
Known for their recurrent structure and the internal mechanism that stores the information on the previous state and forward it to the next state, RNNs are preferred when it comes to sequence-based approaches. There are two types of RNN implementations namely, static and dynamic RNNs. The main difference between them is that dynamic RNN is configured to accept variable length input while static RNN only accepts fixed length input. Having said that, when a static RNN is used, padding or truncation needs to be perform on the input sequence to ensure it matches the defined input length requirement of the model. In contrast, only truncation is needed when for dynamic RNN so that the input sequence length does not exceed the defined required length.

\section{Approach}\label{approach}

\subsection{Overview}\label{approach:overview}
Fig.~\ref{fig:OverviewDetailed} demonstrates the overview of SeqMobile, which contains learning phase and deployment phase.

\noindent{\textbf{Learning phase}}, which is done on a server, consists of \textit{feature preparation, network training, and model conversion}. In the feature preparation step, we focus more on extracting features as sequences such that it can provide more semantics to help distinguish between malicious and benign behaviors inside Android applications. First, a set of feature dictionaries will be constructed (step \textcircled{1}). Next, we extract the feature sequences and use the constructed dictionaries to filter out the redundant elements (step \textcircled{2}).
After that, we represent each element in the filtered sequence with a unique integer identifier and pass it into our proposed network 
for training (step \textcircled{3}).
Once a trained model is obtained, it will be converted into a \textit{TensorFlow Lite}~\cite{tensorflow-lite} model (step \textcircled{4}), which can then be loaded onto mobile device and used with the Tensorflow Lite interpreter to do inferences (step \textcircled{5}). In the midst of conversion, Tensorflow provides user an option to enable quantization, which is a technique that can further reduce the size of pre-trained model with minimal effect on the accuracy. However, we don't perform quantization in SeqMobile since our experiments in \S~\ref{experiment:performance_quantization} shows it is more advantageous for our proposed network.

\noindent{\textbf{Deployment phase}} will first perform feature extraction from the target APK and use the constructed dictionaries from the learning phase to filter out the redundant elements (step \textcircled{A}), when an Android package (APK) is downloaded into the device.
To ensure efficiency, a small but crucial part of the feature extraction module is implemented using native code, where a performance gain can be observed even on lower-end devices. 
After that, the extracted sequence will be fed into the classifier module to determine whether the target application is benign or malicious (step \textcircled{B}). Finally, SeqMobile will output the classification results to the user (step \textcircled{C}).

\subsection{Feature Preparation}\label{approach:feature_preparation}
{In the feature preparation step, SeqMobile mainly focuses on selecting features from the AndroidManifest.xml and DEX file (classes.dex). In order to determine the features used in SeqMobile, we perform a static analysis and select 4 kinds of informative and semantic feature sets that can potentially help to distinguish between malware and benign samples. For each feature set, we rationalize the reason of our selection with concrete examples. Besides, to remove the effect of certain elements that may not be helpful in providing information, we have built dictionaries (step \textcircled{1}) for the purpose of filtering out those elements. {Each element in} the filtered sequence will then be represented as a unique integer value so that it can be fed into our neural network for training. In addition, we also experiment with different combinations of feature sets to determine which combination yields the best accuracy with acceptable extracting performance. Based on the results, we selected 4 feature sets as our input to the neural network. The details of the experiment can be found in \S~\ref{experiment:effectiveness_evaluation_feautre_selection_extraction_dnn:accuracy_feature_dnn}.}

\subsubsection{\textbf{Feature selection}}\label{approach:feature_preparation:selection}
{As a result of the strong dependency between the detection accuracy of learning-based approaches and the coverage of malicious information, the features that are able to represent more semantics in the malicious behaviors will have a higher probability to achieve a better result. Thus, based on this concept, we determine to accept two \textit{non-sequential features}: {Permission}, {Intent filters}, and two \textit{sequential features}: {API call sequence}, {Intent sequence} depended on our analysis against Android malware.}

\begin{itemize}[leftmargin=*]

\item
{$N^{\{Perm\}}$}: {Permissions are defined in the AndroidManifest file. Previous studies~\cite{verma2016permissions, hyunjaekangstatic, androdialysis2017} have shown that majority of the malicious application tend to request dangerous permissions as compared to benign applications. This indicates that including permissions as a part of our feature can potentially help us distinguish malware and benign apps.
}

\begin{table}\scriptsize
\caption{An example of API mapping}
\label{tab:feature_preparation:api_mapping}
\begin{center}
 \begin{tabular}{||l l||} 
 \hline
 \textbf{API} & \textbf{Description}  \\  
 \hline\hline
  $API_0$ & Ljava/net/HttpURLConnection;-\textgreater connect  \\ 
 \hline
  $API_1$ & Ljava/io/BufferedReader;-\textgreater readLine  \\ 
 \hline
  $API_2$ & Landroid/net/Uri;-\textgreater parse  \\ 
 \hline
  $API_3$ & Landroid/content/ContentResolver;-\textgreater query  \\ 
 \hline
\end{tabular}
\end{center}
\end{table}

\item
{$N^{\{Intent\}}$}: {Intent filters are defined inside the AndroidManifest.xml file that specifies what type of intents a component (e.g., Activity) would like to receive. Intent filter makes it possible for other applications to directly start the activity by sending out the defined intent message. An example of how malicious application abuse the intent filter is that they usually listen for the {BOOT\_COMPLETED} intent, which is sent after successfully booting up the mobile device, to start their malicious activities{~\cite{androdialysis2017}}. 
}

\item
{$S^{\{API\}}$}: 
{Representing API calls features in an unordered manner is usually sufficient to provide enough information on the behaviors of an Android application. However, if the API calls are represented in a sequential manner, it can provide us with additional semantics amongst the API calls.}
Two behaviors are defined using sequence of API calls, where $S^{\{API\}}_0$=$\{API_0, API_1, API_2, API_3\}$ and $S^{\{API\}}_1$=$\{API_2, API_3, API_0, API_1\}$. The details of each API can be found in {Table~\ref{tab:feature_preparation:api_mapping}}. The behavior $S^{\{API\}}_0$ represents the action of communicating to the internet followed by reading of SMS inbox messages. This is a typical behavior for instant messaging applications where they try to verify your mobile number. In contrast, $S^{\{API\}}_1$ represents the action of reading of SMS inbox messages followed by a communication to the internet, which is possibly a malicious behavior where an adversary attempts to retrieve inbox messages.

\item
{$S^{\{Intent\}}$}:
Besides picking API call sequence as a part of our feature set, we also accept the sequential intent as a supplemental feature for the API call sequence with the following two reasons.
Firstly, when API call sequence is included as our feature set, the function parameters are not taken into account, thus, generic API calls such as Landroid/content/Intent;-\textgreater init or Landroid/content/Intent;-\textgreater setAction is unable to provide information on the purpose of invocation unless the corresponding \textit{string} parameter is included. Secondly, relying on intent filters is not sufficient as it represents what type of intent the component is looking out for. Oftentimes, intents can be omitted from the manifest file. One typical example is that malware can make use of the {ACTION\_CALL} intent to call premium rate numbers while the user is not looking{~\cite{mouabad}}. In addition, the {ACTION\_CALL} intent does not need to be defined in the manifest file. Thus, we include the \textit{string} parameter of the generic API calls mentioned above in a sequential order and intertwine it with the API call sequence (e.g., $S^{\{API, Intent\}}$=$\{API_1, IN_0, API_2, API_0, IN_2, API_3\}$).
\end{itemize}

\subsubsection{\textbf{Feature dictionary construction}}\label{approach:feature_preparation:dict_construction}
{Since there are certain elements that may not be related to potential malicious behaviors in the raw data (i.e., .xml and .dex files) within each feature set, a set of feature dictionaries are necessary in the feature extraction step. Thus, based on the static analysis result on potential malicious behaviors, a set of feature dictionaries has been constructed (step \textcircled{1}).}

\noindent{\textbf{Permission and intent dictionary.}} {To build the permission and intent dictionary, we refer to the Android source code which are predefined by Google developers to retrieve all the permissions and intent filter values.}

\noindent{\textbf{API call dictionary.}} {To build the API call dictionary, we conduct a data-driven analysis to collect API calls from more than {60,000} real-world applications and pruned the result set by removing self-defined API calls as well as uncommon third-party API calls~\cite{feng2020performancesensitive}.}

{In summary, we constructed 3 feature dictionaries and the total number of vocabulary in each dictionary can be found in {Table~\ref{tab:feature_preparation:dict_size}}.}

\begin{table}\scriptsize
\caption{Vocabulary size of each dictionary}
\label{tab:feature_preparation:dict_size}
\begin{center}
 \begin{tabular}{||l c||} 
 \hline
 \textbf{Dictionary} & \textbf{Vocabulary size}  \\  
 \hline\hline
  Permissions & 324  \\ 
 \hline
  Intents & 262  \\ 
 \hline
  API calls & 2,288  \\ 
 \hline
\end{tabular}
\end{center}
\end{table}

\subsubsection{\textbf{Feature extraction}}\label{approach:feature_preparation:seq_extraction}
{To generate and formalize the selected features (step \textcircled{2}), we first propose a repetitive elements removal method, which can boost the performance of our system with little effect on the accuracy (discussed in \S~\ref{experiment:effectiveness_evaluation_feautre_selection_extraction_dnn:removal}), by reducing the length of feature sequence for the selected sequence-based features. Next, with the shorten feature sequence, we transform it into a numeral sequence such that it will be suitable to be fed into the neural network.}

\noindent{\textbf{Raw string based feature representation.}}
{To extract $S^{\{API\}}$, we disassemble the DEX file and look for instructions starting with the \textit{invoke-*} opcode. At the same time, we also include the $S^{\{Intent\}}$ feature into the sequence by looking out for instructions that starts with the \textit{const-string} opcode. For each of the instructions found, we concatenate them to form a \textit{string} based sequence, $S_d$. An example of such sequence will be $S^{\{API, Intent\}}$=$\{API_0, API_1, IN_1, IN_2, API_3\}$ where $IN_1$ and $IN_2$ are intent values. After that, permissions and intent filters will be extracted and matched against the corresponding dictionaries to construct the non-sequential feature, $N^{\{Perm\}}$ and $N^{\{Intent\}}$. Finally, we concatenate the extracted sequence, $S_d$, together with $N^{\{Perm\}}$ and $N^{\{Intent\}}$ to form the final sequence, \textit{$S_f$}.
Based on the experiment results presented in \S~\ref{experiment:effectiveness_evaluation_feautre_selection_extraction_dnn:accuracy_feature_dnn}, we select 4 features as the final feature set namely, $N^{\{Perm\}}$, $N^{\{Intent\}}$, $S^{\{API\}}$, and $S^{\{Intent\}}$.}

\noindent{\textbf{Repetitive elements removal.}} 
{Oftentimes, there can be repeated elements in a sequence. An example of how this can occur is that a developer may define two different methods, $M_1$ and $M_2$, that execute very similar tasks. We define $API_i$ and $IN_i$ as API calls and intent values which are defined in the dictionaries and $S\_API_i$ as self-defined API calls not found in the API dictionary. Consider $M_1$= $\{IN_0, API_0, S\_API_0\}$ and $M_2$=$\{IN_0, API_0, S\_API_1, API_1\}$. After the extracted sequence is filtered against the respective dictionaries, the resulting sequence is $\{IN_0, API_0, IN_0, API_0, API_1\}$, which is basically a repetition of \{$IN_0, API_0$\} followed by an $API_1$. If we view it from the perspective of a text sentiment analysis problem, having the sentence \qq{this movie is great, this movie is great} does not change the polarity of the sentence, likewise, having repetitive elements in the sequence will have insignificant effects (\S~\ref{experiment:effectiveness_evaluation_feautre_selection_extraction_dnn:removal}) on the polarity (i.e., malicious or benign). Thus, the resulting sequence after removing the repetitive elements will be $\{IN_0, API_0, API_1\}$.}

\noindent{\textbf{Dictionary identifier assignment and integer based sequence representation.}} {In order to transform the string based sequence into an integer sequence, we first assign each dictionary element with a unique integer identifier and store the pair in a look up table, $T_l$. By referring to the look up table, we represent each element in \textit{$S_f$} with their respective integer identifier. After that, we can feed it into the embedding layer of the proposed network {Table \ref{tab:proposed_network}} (step \textcircled{3}).}

\begin{table}\scriptsize
\caption{Network architecture - Bi-LSTM}
\label{tab:proposed_network}
\begin{center}
 \begin{tabular}{||l l||} 
 \hline
 \textbf{Layer} & \textbf{Output Shape}  \\  
 \hline\hline
 Embedding & (None, None, 128)  \\ 
 \hline
  Bidirectional (LSTM) & (None, None, 512)  \\ 
 \hline
  Batch Normalization & (None, None, 512)  \\ 
 \hline
  GlobalMaxPooling1D & (None, 512)  \\ 
 \hline
 Dense (ReLU) & (None, 64)  \\ 
 \hline
 Dense (ReLU) & (None, 32)  \\ 
 \hline
 Dense (softmax) & (None, 2)  \\ 
 \hline
\end{tabular}
\end{center}
\end{table}

\subsection{Deep Learning Model Construction}\label{approach:dl_model_construction}
{To discover the usability of different neural networks for our selected feature sets, we present 6 basic neural networks to train our classifier (i.e., single layer LSTM/GRU, stacked LSTM/GRU, and Bi-LSTM/GRU) (details on our website~\cite{seqmobile}). By comparing the accuracy, we accept the Bi-LSTM, which yields the highest accuracy, as basic network and perform a customization on the architecture to further enhance the accuracy. The architecture of the customized Bi-LSTM network is shown in {Table~\ref{tab:proposed_network}}.}

\subsubsection{\textbf{Sequence padding}}\label{approach:dl_model_construction:seq_padding}
{The length of \textit{$S_f$} varies from one application to another, hence, we pad the input sequence to ensure that the length is consistent with the required input length of the network, \textit{L}. If the length of \textit{$S_f$} is shorter than \textit{L}, we perform post-zero padding to \textit{$S_f$} till it reaches \textit{L}. If the length of \textit{$S_f$} is longer than \textit{L}, we truncate \textit{$S_f$} till \textit{L}.}

\subsubsection{\textbf{Customized deep neural network architecture}}\label{approach:dl_model_construction:architecture}
{To train our malware classifier, we present a customized Bi-LSTM network. As shown in {Table \ref{tab:proposed_network}}, the first layer is an embedding layer. Due to our large vocabulary size, representing our input sequence using one-hot encoding results in a sparse vector, which is not memory efficient during training and prediction. By incorporating an embedding layer, it can help to reduce the dimensionality of our feature vectors where each element of our integer input sequence is represented as a lower dimension fixed sized vector. The second layer is a Bi-LSTM layer, which has the ability to preserve information from both the forward and backward along the sequence, thus allowing the network to understand the contextual information better and results in a more comprehensive learning of the problem. The third layer is a batch normalization layer, it has been shown that incorporating a batch normalization layer to the network can reduce the internal covariate shift~\cite{batchnormalization2015,cooijmans2016recurrent}.
After that, a global max-pooling function is used to capture the most important factor. Then, the output from the global max-pooling function will pass through the 2 fully connected layers with Rectified Linear Unit (ReLU) activation function followed by a fully connected layer with a softmax activation function. Finally, the output from the last fully connected layer determines which class the input sequence belongs to.}

\subsection{Model Conversion and Quantization}\label{approach:model_conversion_quantization}
{In order to make our pre-trained model compatible with mobile devices, we convert the pre-model into a TensorFlow Lite model using TFLiteConverter \cite{tfliteconverter}. We have also conducted experiments and decided not to perform quantization even though it has advantages such as prediction time and model size reduction. By not applying quantization to our model, we are able to achieve dynamic prediction time where the prediction time cost is dependent on the extracted sequence length (details in \S~\ref{experiment:performance_quantization}).}

\subsection{Real Time Detection System}\label{approach:real_time_detection_system}
{Before conducting a real-time detection on device, the feature dictionaries and converted model will be loaded into the feature extraction and classifier modules respectively (step \textcircled{5}). Once a new APK file is received, SeqMobile first performs feature extraction on the target APK with the constructed dictionaries from the learning phase (step \textcircled{A}). To improve the overall performance of SeqMobile, we directly extract the selected features from the binary files and perform repetitive sequence removal such that the sequence length can be reduced. In addition, we also optimized our feature extraction module by incorporating some native code in the implementation. Next, we perform truncation to the input sequence but not padding. Results from our experiments shows that without padding, we can not only achieve the best accuracy, but also a shorter prediction time. Finally, the classifier module will determine from the extracted features whether the Android application is malicious or benign.}

\begin{table}\scriptsize
\caption{Dataset}
\label{tab:dataset_breakdown}
\begin{center}
 \begin{tabular}{||l c c||} 
 \hline
 \textbf{Class} & \textbf{Source} & \textbf{Quantity}  \\  
 \hline\hline
 Malicious & Contagio & 327  \\ 
 \hline
  Malicious & Drebin & 5,339 \\ 
 \hline
  Malicious & Genome & 1,253 \\ 
 \hline
  Malicious & Pwnzen & 774 \\ 
 \hline
  Malicious & Virusshare & 14,797 \\ 
 \hline
  Benign & Google Play Store & 22,490  \\
 \hline
 \hline
 Total Malicious & - & 22,490 \\
 \hline
 Total Benign & - & 22,490 \\
 \hline
 \textbf{Total} & - & 44,980 \\ 
 \hline
\end{tabular}
\end{center}
\end{table}

\begin{table*}[t]\scriptsize
\caption{Detection results of best feature category combinations across difference networks}
\label{tab:experiment:bestfeaturediffnetwork}
\begin{center}
\begin{threeparttable}
 \begin{tabular}{||l l c c c c||} 
 \hline
 \textbf{Network} & \textbf{Feature Combination} &\textbf{Sequence Length} & \textbf{Accuracy} & \textbf{Precision} & \textbf{Recall}    \\  
 \hline\hline
 \begin{tabular}[c]{@{}c@{}}Single LSTM\end{tabular}  & $N^{\{Perm\}}$, $N^{\{Intent\}}$, and $S^{\{API\}}$ & 600 & 96.47\% & 97.66\% & 95.23\%\\ \hline
 \begin{tabular}[c]{@{}c@{}}Single GRU\end{tabular}  & $N^{\{Perm\}}$, $N^{\{Intent\}}$, and $S^{\{API\}}$ & 1,500 & 96.78\% & 96.99\% & 96.56\%\\ \hline
 \begin{tabular}[c]{@{}c@{}}Stacked LSTM\end{tabular} & $N^{\{Perm\}}$, $N^{\{Intent\}}$, and $S^{\{API\}}$ & 600 & 96.49\% & 97.86\% & 95.05\%\\\hline
 \begin{tabular}[c]{@{}c@{}}Stacked GRU\end{tabular} & $N^{\{Perm\}}$, $N^{\{Intent\}}$, and $S^{\{API\}}$ & 1,500 & 96.90\% & 97.42\% & 96.35\%\\\hline\rowcolor{gray!30}
 \begin{tabular}[c]{@{}c@{}}Basic Bi-LSTM\end{tabular} & $N^{\{Perm\}}$, $N^{\{Intent\}}$, and $S^{\{API\}}$ & 1,500 & 96.98\% & 97.51\% & 96.41\%\\ \hline
 \begin{tabular}[c]{@{}c@{}}Basic Bi-GRU\end{tabular} & $N^{\{Perm\}}$ and $S^{\{API, Intent\}}$ & 1,900 & 96.75\% & 97.47\% & 96.00\%\\\hline\hline\rowcolor{gray!30} 
 \begin{tabular}[c]{@{}c@{}}Customized Bi-LSTM\end{tabular} & $N^{\{Perm\}}$, $N^{\{Intent\}}$, and $S^{\{API, Intent\}}$ & 1,700 & 97.85\% & 97.89\% & 97.81\%\\ \hline
 \begin{tabular}[c]{@{}c@{}}Customized Bi-GRU\end{tabular} & $S^{\{API\}}$ & 1,500 & 97.67\% & 98.11\% & 97.21\%\\ \hline
 \hline
\end{tabular}
\end{threeparttable}
\end{center}
\end{table*}

\section{Experiments}\label{experiment}
{In this section, we present four sets of experimental studies. We aim to determine: (1) the detection performance of different networks across different combination of feature categories; (2) the training and prediction performance gain through our repetitive elements removal method; (3) the feature extraction performance across different mobile devices; and (4) the performance comparison between the quantized and non-quantized dynamic RNN model. Finally, we briefly compare the performance of our approach with two other previous work~\cite{feng2019performance,feng2020performancesensitive}.}

\subsection{Experiment Environment and Dataset}
\noindent \textbf{Environment}. 
{All experiments are conducted on an Ubuntu server with Intel Xeon E5-2699 V3 CPUs, NVIDIA GeForce RTX 2080 Ti GPU and 6 different Android devices, which consists of 3 flagship devices (i.e., Samsung Note10+, S10+, S9+), 1 common device (i.e., Samsung S7), and 2 low-end devices (i.e., Samsung J2 Pro and HTC ONE A9).}
 
{In our server-end tasks, Java is our choice of language for implementing the feature extraction module and TensorFlow 2{~\cite{tensorflow}} is chosen as our deep learning framework. For feature extraction, we use additional tools such as \textit{dexdump}~{~\cite{dexdump2018}} 
and AXMLPrinter2\cite{AXMLPrinter2}. As for the TensorFlow Lite converter, a specific TensorFlow nightly build (2.2.0.dev20200430)~\cite{tfnightly} which supports our proposed network architecture is used.}

\noindent \textbf{Dataset}.
{To evaluate SeqMobile, we collected 44,980 Android applications samples which can be divided into two classes; benign and malicious. We crawled the benign samples from Google Play Store, while the malicious set is composed of samples from different sources such as Drebin\cite{arp2014drebin}, Genome project\cite{zhou2012dissecting}, Contagio Mobile\cite{contagio}, VirusShare\cite{virusshare}, and Pwnzen Infotech Inc.~\cite{chen2016stormdroid, chen2018automated}. The breakdown of the dataset is shown in {Table~\ref{tab:dataset_breakdown}}. To split our dataset, we randomly select 70\% of the samples from each class for training, 15\% for validation and 15\% for testing.}

\begin{figure}[t]
\centering
\includegraphics[scale=0.2]{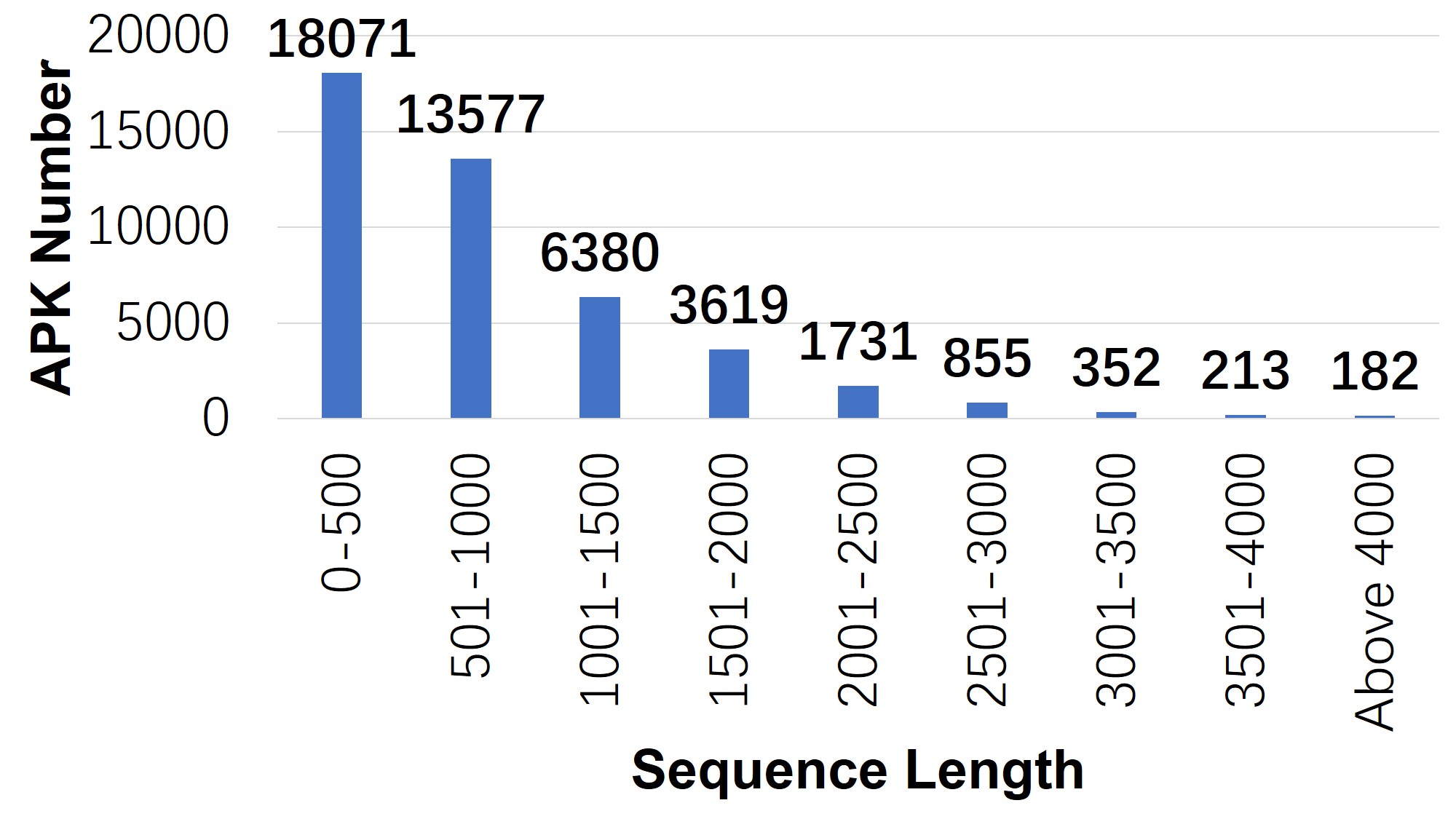}
\caption{{Distribution of extracted sequence length}}
\label{fig:APK_SEQLEN_FREQ}
\end{figure}

\subsection{Effectiveness Evaluation of Feature Selection, Deep Neural Networks, and Repetitive Pattern Removal in Feature Preparation and Network Training Phases}
\label{experiment:effectiveness_evaluation_feautre_selection_extraction_dnn}
{We evaluate the effectiveness of SeqMobile in the learning phase from the following aspects: (1) the accuracy across different feature categories, length, and deep neural networks and (2) the effect on training accuracy and performance brought by our repetitive pattern removal method.}

\begin{figure}[t]
\centering
\includegraphics[scale=0.2]{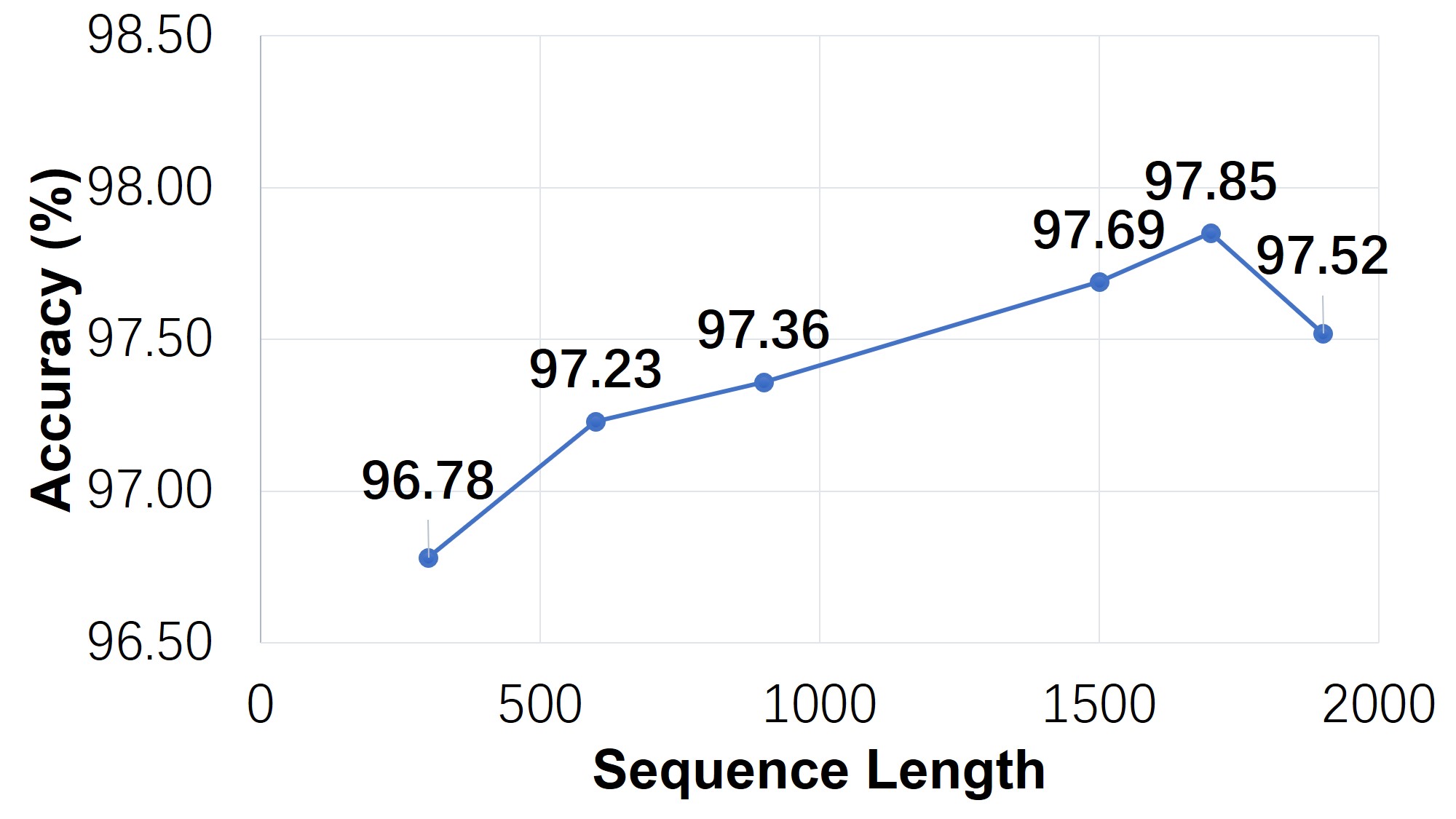}
\caption{Accuracy change across different sequence length on $N^{\{Perm\}}$, $N^{\{Intent\}}$, and $S^{\{API, Intent\}}$}
\label{fig:F9_TFLite_Accuracy}
\end{figure}

\subsubsection{Accuracy comparison across feature categories and deep neural networks}\label{experiment:effectiveness_evaluation_feautre_selection_extraction_dnn:accuracy_feature_dnn}
{To determine the best training configuration, we set up an experiment across three aspects (i.e., feature categories, sequence length, and network types).
To find out the most suitable network, we first evaluate the detection accuracy of 6 different basic network configurations across {8} different feature categories (details on our website~\cite{seqmobile}), which apply $S^{\{API\}}$ as the basic feature and combined with $N^{\{Perm\}}$, $N^{\{Intent\}}$, and $S^{\{Intent\}}$ respectively. In addition, as shown in {Fig.~\ref{fig:APK_SEQLEN_FREQ}}, majority of the APKs in our dataset have sequence length ranging up to 2,000. Thus, we also train each network across different sequence length to determine the appropriate sequence length that yields the best accuracy. To choose the range of sequence length to experiment on, we progressively increase the sequence length until no significant improvement on the accuracy can be observed. In total, according to the distribution of extracted sequence length, we select 6 different values ranging from 300 to 1,900 to train the networks.
As shown in {Table~\ref{tab:experiment:bestfeaturediffnetwork}}, the basic Bi-LSTM network achieves the highest accuracy (i.e., 96.98\%) out of the other basic networks. 
To further improve the detection accuracy, we customize the basic Bi-LSTM to our best effort such that there is a significant improvement on the accuracy. We first progressively increase the number of parameters (i.e., increase LSTM units and add fully connected layer) in the network. Consequently, a longer training duration can be observed due to the increased network parameters. Thus, we added a batch normalization followed by \textit{GlobalMaxPooling1D} layer to reduce the training duration. Our experiment results shows that by configuring the network as shown in {Table \ref{tab:proposed_network}}, the feature set combination that achieves the best accuracy is $N^{\{Perm\}}$, $N^{\{Intent\}}$, and $S^{\{API,Intent\}}$, which achieves an accuracy of 97.85\% when the sequence length is 1,700 (shown in {Fig.~\ref{fig:F9_TFLite_Accuracy}}).}

\subsubsection{Effect of repetitive pattern removal}\label{experiment:effectiveness_evaluation_feautre_selection_extraction_dnn:removal}
{We conduct experiments to investigate how removing of repetitive sequence will benefit the performance of the system. In the experiment, we use the following metrics: (1) accuracy and (2) training duration to evaluate our method. The results show that (1) removing the repetitive pattern in a sequence has insignificant effects on the learning ability of the neural network; (2) it can help improve the training and prediction performance in terms of time cost.}

\begin{table}\scriptsize
\caption{Average sequence length}
\label{tab:avg_elements_removed}
\begin{center}
 \begin{tabular}{||l c c||} 
 \hline
 \textbf{Dataset} & \textbf{Original Length} &\textbf{After Removal} \\  
 \hline\hline
 Benign & 2,789 & 1,060  \\ 
 \hline
  Malware & 1,249 & 598  \\ 
 \hline
\end{tabular}
\end{center}
\end{table}

\noindent\textbf{Average number of removed repetitive elements.}
{To show the necessity of our repetitive elements removal method, we first calculate the number of repetitive elements removed in each dataset (i.e., benign and malware) to get a rough estimation of the proportion being removed by comparing with the original average sequence length. 
As shown in {Table~\ref{tab:avg_elements_removed}}, the average sequence length is reduced by approximately 62\% (1,060 vs. 2,789) in the benign dataset. Similarly, the average sequence length is reduced by approximately 52\% for the malware dataset. When taking into account for both datasets, the sequence length is reduced by 57\% on average. Thus, based on the results, we take 0.6 as an approximate proportion of the repetitive elements in the original sequences and define equation (\ref{eq:original_seq_len}) to calculate the estimated original sequence length for each non-repetitive sequence length.}

\begin{equation}
\tvar{Seq_{original}}=
\frac{\tvar{Seq}_{non-repetitive}}{1-0.6}
\label{eq:original_seq_len}
\end{equation}

\begin{figure}[]
\centering
\includegraphics[scale=0.24]{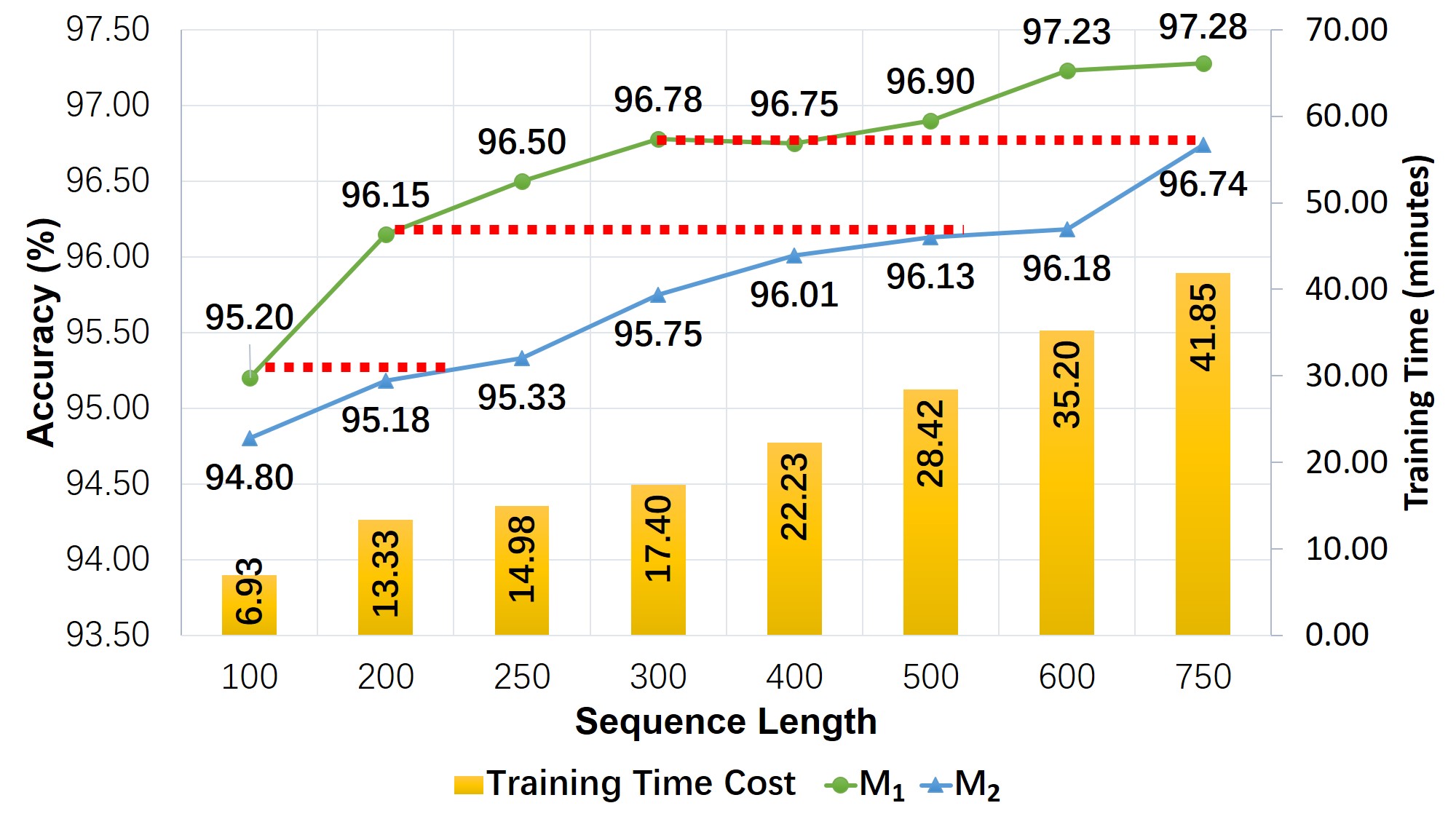}
\caption{Training time across different sequence length and accuracy comparison between $M_1$ and $M_2$}
\label{fig:SeqLenMapping}
\end{figure}

\noindent\textbf{Accuracy comparison between non-repetitive and original sequence.}
To discover the effect of our repetitive elements removal method on model accuracy, we define the following 2 models, $M_1$, which adopts the sequences without repetitive elements as its input, $Seq_{non-repetitive}$, and $M_2$, which is trained using the original sequences that contain repetitive elements, $Seq_{original}$, and train them with sequence length ranging from 100 to 750. 
From the previous experiment, the input sequence length is reduced by 57\% on average after removing repetitive elements, as a result, with the same sequence length, $Seq_{non-repetitive}$ will provide more information than $Seq_{original}$. {To compare the accuracy of $M_1$ and $M_2$, we use the equation in (\ref{eq:original_seq_len}) to estimate the corresponding original sequence length for each $Seq_{non-repetitive}$, and assign them as a control group. For example, if the accuracy of $M_1$ at sequence length 200 is 96.15\%, we will apply equation (\ref{eq:original_seq_len}) to calculate the length of $Seq_{original}$ (i.e., 500). From the line chart in {Fig.~\ref{fig:SeqLenMapping}}, the accuracy for $M_2$ is 96.13\% when the sequence length is 500, which is very close to the accuracy for $M_1$ (i.e., 96.15\%) at sequence length 200. Similarly, the accuracy at $M_1$ when the sequence length is 300, is very close to the accuracy of $M_2$ when the sequence length is at 750 (96.78\% vs. 96.74\%). This shows that by removing repetitive elements in a sequence, it will have very little effect on the learning ability of the network.}

\noindent\textbf{Performance improvement in network training.}
{To find out the performance improvement of our repetitive elements removal method in the network training phase, we also provide a comparison on training time between each grouped models with non-repetitive and original sequences respectively. From the histogram in {Fig.~\ref{fig:SeqLenMapping}}, the training time for $M_2$ is 28.42 minutes when the sequence length is 500. Comparing with the time for $M_1$ (i.e., 13.33 minutes) at sequence length 200, our repetitive elements removal method improves the training performance by around 53.1\%. Similarly, comparing the training time for $M_1$ when the sequence length is 300, we observe a much shorter training time for $M_2$ when the sequence length is at 750 (17.40 minutes vs. 41.85 minutes). This provides a strong evidence for proving that our repetitive elements removal method benefits a lot in the training performance.}

\subsection{Performance Evaluation and Optimization of Feature Extraction and Detection on Android Mobile Devices}\label{experiment:performance_evaluation_optimization_feature_extraction_detection_mobile}
{We evaluate the performance of our device-end modules (i.e., feature extraction and prediction modules) separately. For the feature extraction module, we conduct experiments to determine (1) the performance of extraction time on mobile devices, by comparing the extraction time of our final feature combination (i.e., $N^{\{Perm\}}$, $N^{\{Intent\}}$ and $S^{\{API, Intent\}}$) across 6 different APK sizes; and further evaluate (2) the performance gain of the feature extraction module where certain parts are rewritten in native code. For the prediction module, we assess the performance gain across 2 aspects, which are input sequence and deployed model configuration.}

\subsubsection{Performance comparison of feature extraction between implementing with JNI and Java language on Android devices}\label{experiment:performance_evaluation_optimization_feature_extraction_detection_mobile:performance_feature_extraction_jni}
{To discover the potential performance optimizations in feature extraction on real devices, we implement part of the feature extraction module that involves \textit{string} manipulation in C++ and use JNI to interact with the native implementation. Based on the previous experiment, we select the feature set combination that yields the highest accuracy (i.e., $N^{\{Perm\}}$, $N^{\{Intent\}}$ and $S^{\{API, Intent\}}$) and measure the time cost between the Java and JNI implementation. The APKs used in this experiment are handpicked, with file sizes ranging from 5MB to 50MB. We then calculate the performance gain between the implementations using equation (\ref{eq:percentage_error}).}
\begin{equation}
\tvar{Performance\, gain}=
(1-\frac{\tvar{Timecost}_{JNI}}{\tvar{Timecost}_{Java}})\times100\%
\label{eq:percentage_error}
\end{equation}

\begin{figure}[t]
\centering
\includegraphics[scale=0.18]{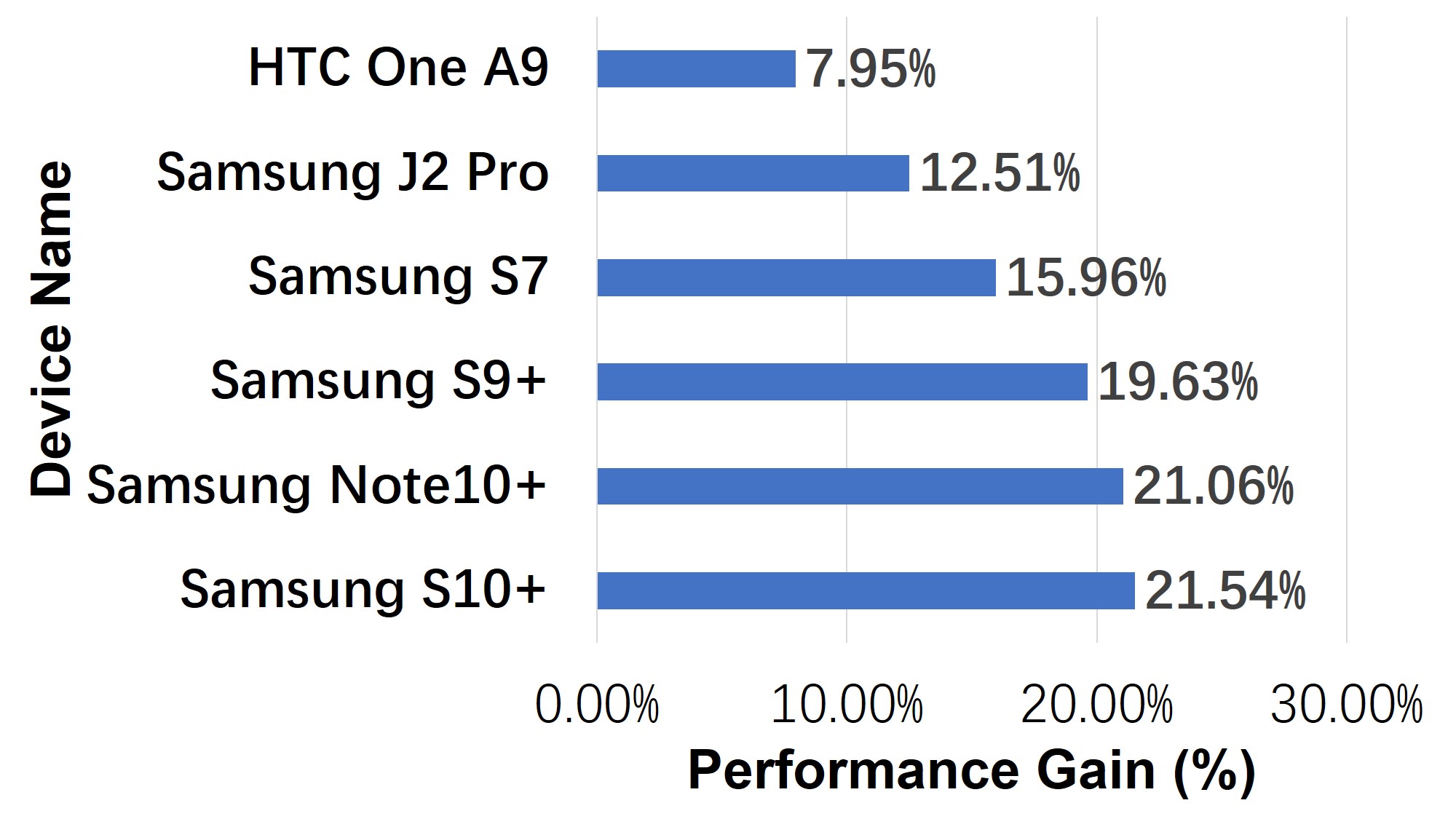}
\caption{Performance gain with JNI implementation}
\label{fig:JNI_PERFORMANCE_INCREASE}
\end{figure}

{As shown in {Fig.~\ref{fig:JNI_PERFORMANCE_INCREASE}}, the JNI implementation can improve the performance of the feature extraction module by approximately 21\% on flagship phones. Even on low-end devices such as HTC ONE A9, a 7.95\% increase in performance can be observed.} 

\begin{figure}[t]
\centering
\includegraphics[scale=0.22]{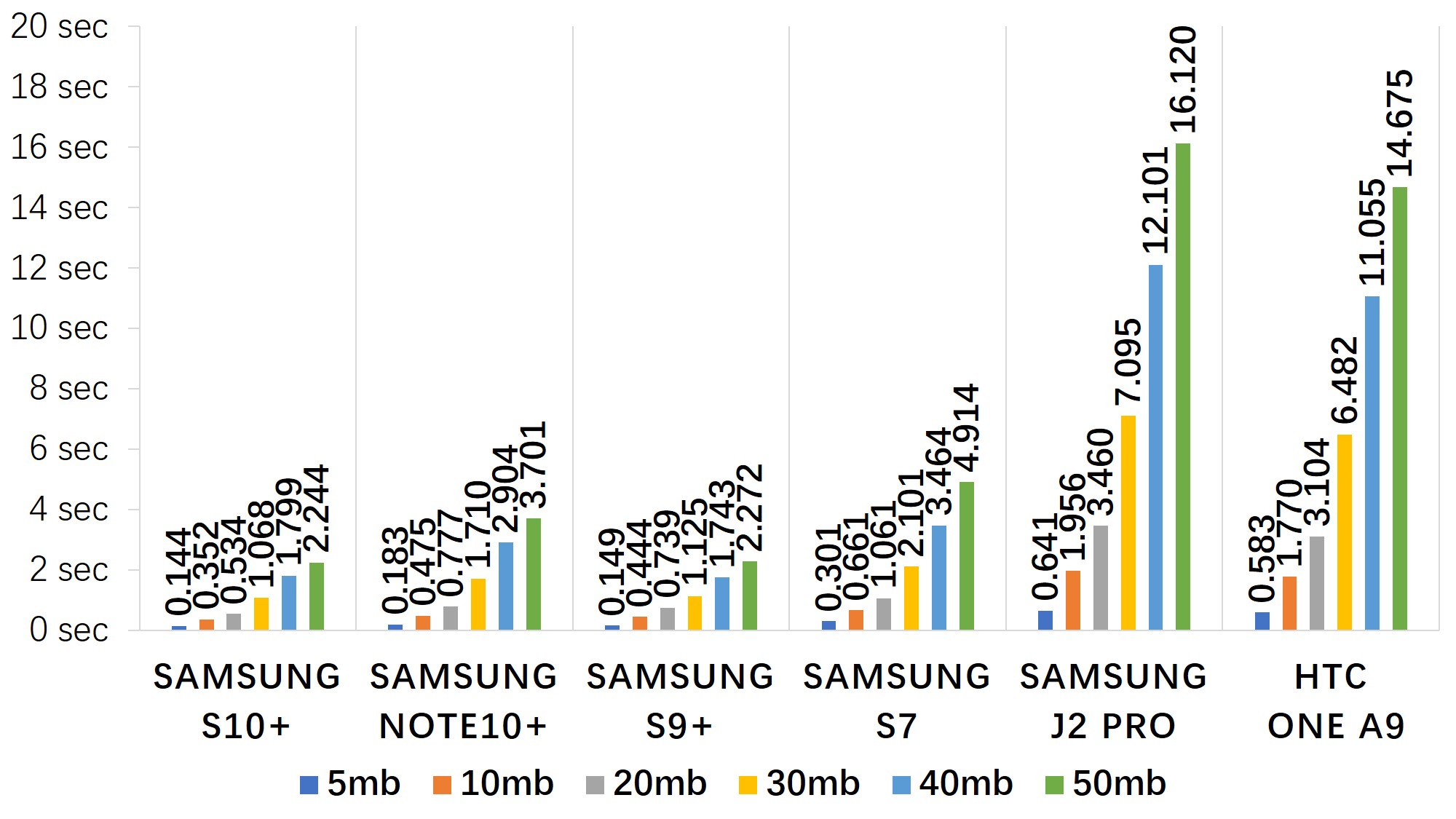}
\caption{Feature extraction (JNI implementation) time cost for different APK sizes across different devices}
\label{fig:JNI_FE_F9_DEVICES}
\end{figure}

{A detailed chart of the feature extraction (JNI implementation) time cost across different devices is shown in {Fig.~\ref{fig:JNI_FE_F9_DEVICES}}. We observe that flagship phones such as Samsung S10+, Samsung Note10+ and Samsung S9+, are able to extract the features faster than common and lower end devices (Samsung S7, Samsung J2 Pro, and HTC ONE A9). The time cost to extract 5MB applications on the flagship phones is between 0.144s and 0.183s. While on common and low-end devices, it takes approximately 0.301s to 0.641s to extract the features. Similarly, for 50MB applications, the flagship phones outperform the low-end devices (3.701s vs. 16.120s).}

\subsubsection{Performance optimizations in prediction on Android devices}\label{experiment:performance_evaluation_optimization_feature_extraction_detection_mobile:performance_optimization_prediction}
{To provide a guidance for improving the performance of sequence-based learning approaches, which may also accept RNN as their computational layer, in the real-time prediction on mobile devices, we conduct two experiments from different aspects. (1) We investigate the performance optimization, which brought by our proposed repetitive pattern removal method, by comparing the prediction time between the original and non-repetitive sequence inputs. (2) We evaluate the state-of-the-art model optimization toolkit provided by TensorFlow for our sequence-based approach. In the second experiment, we first conduct a preliminary investigation to determine the characteristic differences (i.e., model size, input requirements, etc.) between the quantized and non-quantized models. Based on the findings, we further investigate the influence of quantization by comparing the time taken to predict sequences of different lengths between the quantized and non-quantized dynamic RNN models.}

\begin{figure}[]
\centering
\includegraphics[scale=0.22]{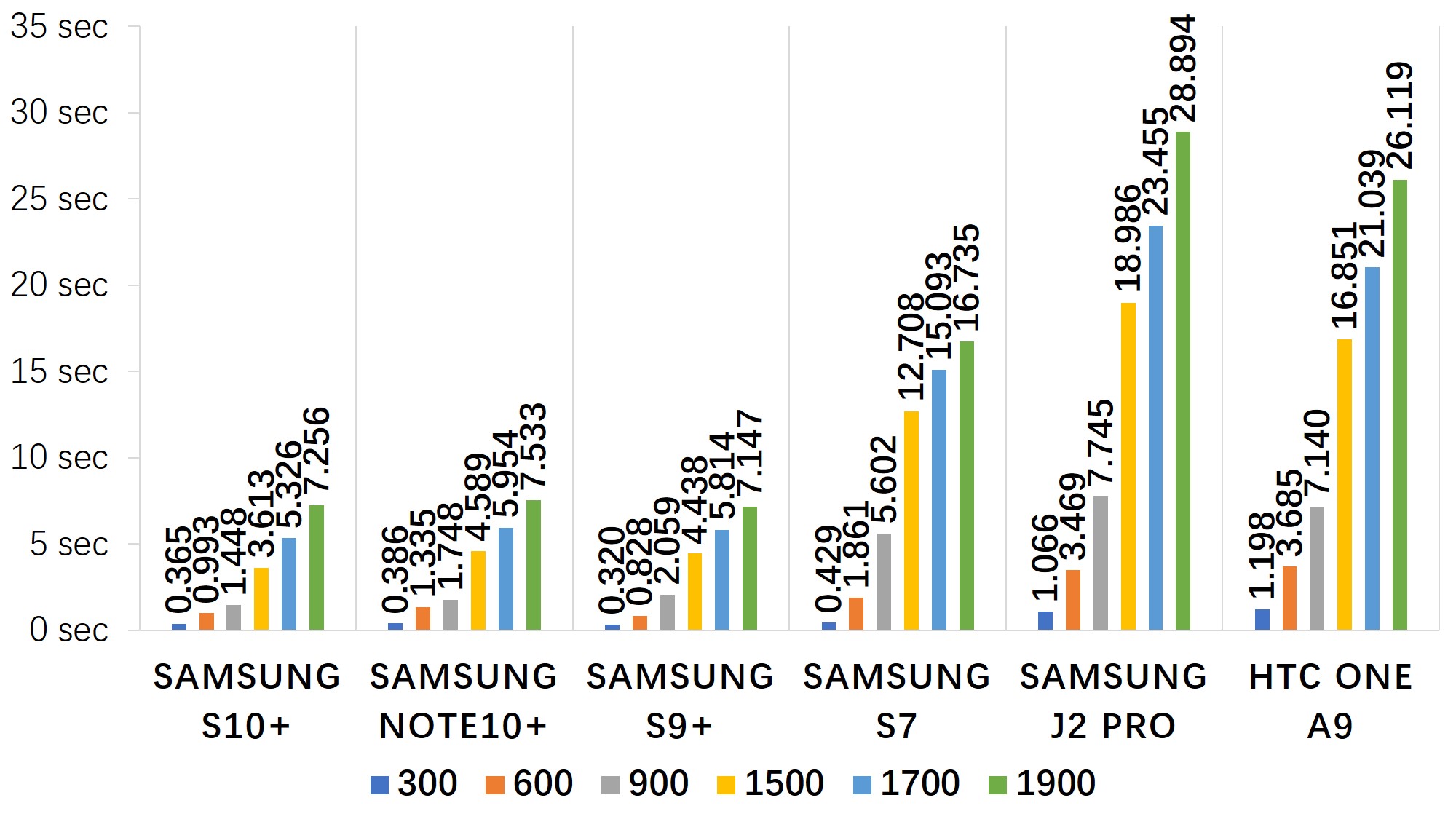}
\caption{Prediction time cost for different sequence lengths across different devices}
\label{fig:F9_PREDICT_TIMECOST_DEVICES}
\end{figure}

\noindent\textbf{Repetitive pattern removal.}
{To discover the effect of our proposed repetitive element removal method in prediction on real devices, we conduct an experiment to measure the prediction time for 6 different sequence lengths across 6 different devices. In this experiment, we randomly pick 5 APKs from each sequence length category (i.e., 300, 600, 900, 1,500, 1,700, and 1,900).}

{As shown in {Fig.~\ref{fig:F9_PREDICT_TIMECOST_DEVICES}}, flagship devices such as Samsung S10+ takes approximately 2.72 times longer (0.365s vs. 0.993s) to predict a sequence that is double the original length (e.g., 300 vs. 600). Similarly for low-end devices such as Samsung J2 Pro and HTC ONE A9, the prediction time is approximately 3.25 times longer (1.198s vs. 3.685s) to predict a sequence that is double the original length. We observe that it takes at least a twofold increase in prediction time cost to predict a sequence that is twice the original length. Based on our experiment in \S~\ref{experiment:effectiveness_evaluation_feautre_selection_extraction_dnn:removal}, we observe that our method reduces the sequence length by approximately 57\%, which directly translates to an improvement in prediction time of at least twofold. Apart from the removal method, different from the traditional sequence-based learning approaches that work on server end, approaches on real devices, which may have a strong performance limitation, should take the input sequence length of their defined networks as an important factor to optimize their real-time performance on device.}

\begin{table}\scriptsize
\caption{Non-quantized and quantized model size comparison}
\label{tab:experiment:quantized_model_size}
\begin{center}
\begin{threeparttable}
 \begin{tabular}{||c c c||} 
 \hline
 \textbf{Seq Len.} & \textbf{Non-quantized(MB)} & \textbf{Quantized(MB)}\\ 
 \hline\hline
 300  & 7.62 & 2.97\\ 
 \hline
  600  & 7.62  & 3.12\\ 
 \hline
  900  & 7.62  & 3.27\\ 
 \hline
  1,500  & 7.62  & 3.56\\ 
 \hline
 1,700  & 7.62  & 3.66\\ 
 \hline
 1,900  & 7.62  & 3.75\\ 
 \hline
\end{tabular}
\end{threeparttable}
\end{center}
\end{table}

\noindent\textbf{Model quantization.}
{To investigate the potential performance oriented influence and their corresponding factors of quantization in our proposed system, we first conduct a preliminary investigation to figure out the main differences between a quantized and non-quantized model. In this experiment, we first compare the model size of the quantized and non-quantized models across different sequence length. As shown in {Table~\ref{tab:experiment:quantized_model_size}}, when the sequence length is 300, the model size is reduced by 61\% (7.62MB to 2.97MB). However, as the sequence length increases to 1,900, the model size is reduced by approximately 50.8\% (7.62MB to 3.75MB). From the results, it is evident that the size of the quantized model is dependent on the sequence length (i.e., longer sequence length will constitute to a larger file size), while the size for the non-quantized model remains constant across different sequence length.}

{Next, we also conduct experiments to compare the accuracy between the quantized and non-quantized model against the pre-trained model. The accuracy of quantized and non-quantized model remains 97.85\%, which is same as the pre-trained model.
{Otherwise, we also notice that if the input requirements of a network is designed to allow variable length input (e.g., dynamic RNN), quantizing the model will remove the flexibility for allowing variable length input (i.e., inputs will be static fixed length). On the contrary, the input requirements remain unchanged if quantization is not applied. In the event where a quantized model is deployed, padding or truncation is required to ensure that the input sequence is of a certain length.}}

\begin{figure}
\centering
\includegraphics[scale=0.24]{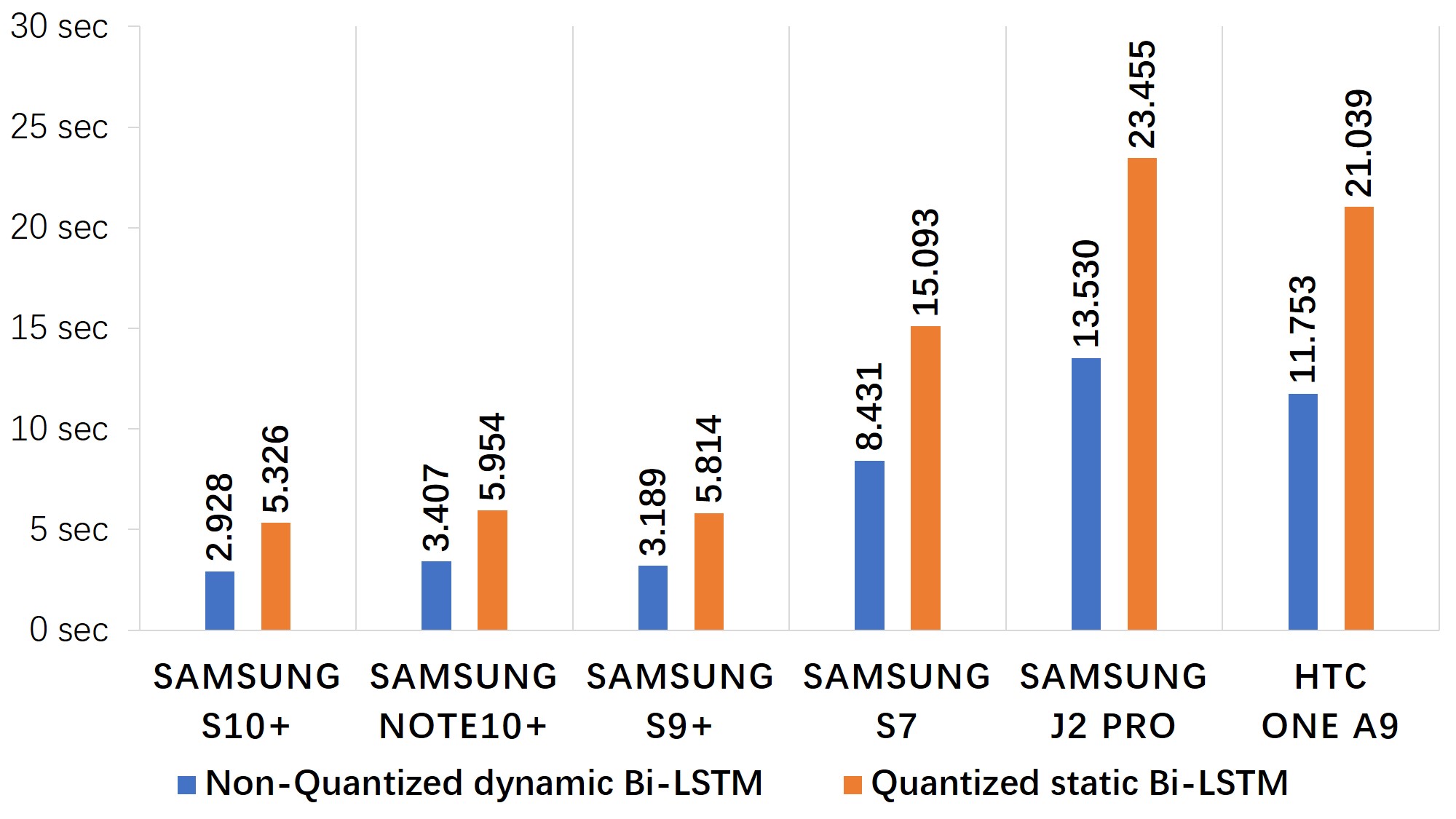}
\caption{Average prediction time between non-quantized dynamic Bi-LSTM and quantized static Bi-LSTM across different devices }
\label{fig:PREDICT_TIME_COST_Q_vs_NQ}
\end{figure}

\noindent\textbf{Dynamic RNN.}\label{experiment:performance_quantization}
{Since our dataset consists of sequences of varying lengths, designing our network to use dynamic RNN will be beneficial in the deployment phase. Based on the findings from the preliminary investigation, we observe that quantizing our model is equivalent to using a static RNN model, hence, we refer to the quantized model as quantized static Bi-LSTM model and non-quantized dynamic Bi-LSTM for the non-quantized model.}

{We conduct an experiment to determine if quantizing our dynamic RNN will improve the performance of our system. In this experiment, we apply the same test set from our previous experiment, which contains 30 APKs with 6 different sequence length. Next, we accept the pre-trained model with the highest accuracy to compare the average prediction time between the quantized static Bi-LSTM and non-quantized dynamic Bi-LSTM models. As shown in {Fig.~\ref{fig:PREDICT_TIME_COST_Q_vs_NQ}}, we observe that the average prediction time for the non-quantized dynamic Bi-LSTM model is much faster than the quantized static Bi-LSTM model. On flagship phones such as Samsung S10+, the average time taken to predict an extracted sequence is 2.928s when using the non-quantized dynamic Bi-LSTM model. On the contrary, it takes 5.326s on average to predict with the quantized static Bi-LSTM model. The average time cost is lowered by approximately 45\% across all devices. The root cause of this observation is expected as padding or truncation is required for the quantized model to make the input length consistent, i.e., 1,700, which constitutes to a higher average prediction time. Based on the results, we decide not to quantize our pre-trained model as it brings about more advantages for our proposed network. The allow for variable length input in the non-quantized dynamic Bi-LSTM model has enable us to achieve a dynamic prediction time as no padding is required (i.e., the time cost per prediction is dependent on the extracted sequence length). On the contrary, a fixed length input is required for the quantized static Bi-LSTM model, where each sequence is padded or truncated to a certain length. By doing so, it constitutes to a consistently higher time cost per prediction. Although the model size is reduced by half (7.62MB to 3.66MB) after quantization, compromising time cost for model size is not a feasible option for our performance-sensitive malware detection system. Hence, for any sequence based performance-sensitive learning approaches, which uses RNN as the basic computational layer, dynamic RNN is currently an important option to optimize their system performance instead of quantization, which is widely adopted.}

	Currently, TensorFlow does not support quantization for models that accept variable length input (e.g., dynamic RNN)\cite{quantizeddynamiclength}. During quantization, a \qq{reshape} operation is added internally to ensure that the input requirement is of fixed length. Thus, padding or truncation of the sequence is required, which results in the dynamic RNN model being indirectly converted into a static RNN model.

\subsection{Comparison between previous work and SeqMobile}
{We briefly compare SeqMobile against two other previous works (i.e., MobiDroid~\cite{feng2019performance} and MobiTive~\cite{feng2020performancesensitive}). We use our results for Samsung S10+ as a baseline to benchmark against Nexus 6P from MobiDroid and Huawei P30 from MobiTive.
As shown in {Table~\ref{tab:compare_prev_work}}, even though the features used in the other two systems are similar (e.g., API calls and manifest properties), our sequence-based approach, which contains additional semantics, is able to achieve a much higher accuracy (i.e., 97.85\%). Although there is an improvement in time cost when comparing to MobiDroid (4.16s vs. 17.76s), our time cost is still approximately 10 times higher (4.16s vs. 0.46s) when compare to MobiTive. Despite our high time cost, with the additional semantics from the sequence-based features, our accuracy is higher than MobiTive (96.75\% vs. 97.85\%). Also, the study from MobiTive~\cite{feng2020performancesensitive} shows a trend of mobile hardware performance improving over the years, we strongly believe that our approach can achieve a significantly lower time cost in the upcoming years, making the accuracy of the detection system a much more important factor.}

\begin{table}\scriptsize
\caption{Comparison of SeqMobile against previous work}
\label{tab:compare_prev_work}
\begin{center}
 \begin{tabular}{||c c c c||} 
 \hline
 \textbf{Systems} & \textbf{Features used} &\textbf{Accuracy(\%)} &\textbf{Time cost(s)} \\  
 \hline\hline
  MobiDroid & \makecell{Opcode sequence\\ API calls\\ 3 Manifest properties}  & 97.35\% & 17.76\\
 \hline
  {MobiTive} & \makecell{API calls\\ 3 Manifest properties} & 96.75\% & 0.46 \\
 \hline
  SeqMobile & \makecell{API and intent sequence\\2 Manifest properties\\} & 97.85\% & 4.16\\
 \hline
 
\end{tabular}
\end{center}
\end{table}

\section{Related Work}\label{related_work}
{In this section, an overview of the current deep learning malware detection approaches will be presented. Generally, these approaches can be categorized into server end and device end approaches.}

\subsection{Server-end Approach}
{Malware detection on server end are usually efficient due to the computational resources available. To summarize, there are approaches~\cite{apkauditor, aung2013permission, Rovelli2014PMDSPM, vinayakumar,chen2016towards,fan2016poster} that extract features (e.g., permissions and API calls) from XML and DEX files and represent them as feature vectors to uncover potential malicious behaviors. There are also approaches~\cite{mclaughlin2017deep, robinnix2017, xiao2017androidmd, opcodeseq2014} that represent features (e.g., API calls, opcode, system calls) as sequences to detect malware. Also, some approaches~\cite{zhuoma2019, ALAM2017230, cdgdroid2018, Yan2019ClassifyingMR} use the additional computation resources to their advantage and analyze complex features such as control flow graphs. While server end approaches achieve great success in detecting malware, it also incurs network transmission overhead for sending the required files to the server.}

\subsection{Device-end Approach}
{Device-end solutions are often performance-sensitive systems that can effectively detect malware under a certain time constraint. Given the performance limitations of Android devices, such device-end solutions usually have limited resources to extract and analyze complex features to classify the applications. Thus, only limited features can be used to detect malware. Feng et al. proposed MobiDroid~\cite{feng2019performance}, a performance-sensitive malware detection system, which represents features (e.g., opcode sequence, API calls, and manifest properties) as feature vectors and leverages deep learning algorithms to help detect malware. Another study by Feng et al.~\cite{feng2020performancesensitive}, uses a different method to directly extract features (e.g., API calls and manifest properties) from binary files to perform classification with a low overhead time.}
This paper studies the device-end performance of sequence-based malware detection and propose performance optimization methods for sequence-based learning approaches.
Closest to our device-end sequence-based approach, Elmouatez et al. proposed Maldozer~\cite{maldozer}, which uses API sequences to detect Android malware families. Differently, they did not consider the performance on devices as a first order factor in the approach.
{Existing device-end approaches mainly focus on extracting limited feature sets to meet the time constraints. With the rapid advancement in technology, the performance of Android devices is improving every year~\cite{feng2020performancesensitive}. Thus, studies on device-end malware detection systems is much needed.}

\section{Conclusion}\label{conclusion_future_work}
{In this paper, we have investigated the effectiveness of using sequence-based learning approach together with performance optimization methods to detect malicious applications on device end. The evaluation results show that our approach achieves a high accuracy (97.85\%) and a reasonable detection time on flagship phones. Moreover, we have also provided a guidance on the state-of-the-art TensorFlow model optimization toolkit for device-end sequence-based approaches.} 

\section*{Acknowledgments}
This work was supported by Singapore Ministry of Education Academic Research Fund Tier 1 (Award No. 2018-T1-002-069), the National Research Foundation, Prime Ministers Office, Singapore under its National Cybersecurity R\&D Program (Award No. NRF2018 NCR-NCR005-0001), the Singapore National Research Foundation under NCR Award Number NSOE003-0001, NRF Investigatorship NRFI06-2020-0022, the National Research Foundation, Prime Ministers Office, Singapore under NCR Award Number NRF2018NCR-NSOE004-0001, the National Natural Science Foundation of China (No. 61902395). We gratefully acknowledge the support of NVIDIA AI Tech Center (NVAITC).

\ifCLASSOPTIONcaptionsoff
\fi

\balance
\bibliographystyle{IEEEtran}
\bibliography{ref}
\end{document}